\documentclass[aps,pra,a4paper,twocolumn]{revtex4}
\usepackage{graphicx}
\usepackage[dvips]{epsfig}
\usepackage{subfigure}
\usepackage{bm}
\begin{document}
\title{Vibrational energy transfer in ultracold
molecule - molecule collisions}

\author{Goulven Qu\'{e}m\'{e}ner and Naduvalath Balakrishnan}
\affiliation{Department of Chemistry, University of Nevada Las Vegas,
Las Vegas, NV 89154, USA}

\author{Roman V. Krems}
\affiliation{Department of Chemistry, University of British Columbia, Vancouver, B.C. V6T 1Z1, Canada}

\date{\today}

\begin{abstract}
We present a rigorous study of vibrational relaxation in  
p-H$_2$ + p-H$_2$ collisions at cold 
and ultracold temperatures and identify an efficient mechanism of ro-vibrational energy transfer.  
If the colliding molecules are in different rotational and vibrational levels, the internal energy 
may be transferred between the molecules through an extremely state-selective process
involving simultaneous conservation of internal energy and total rotational angular momentum.
The same transition in collisions of distinguishable molecules
corresponds
to the rotational energy transfer from one vibrational state of the colliding molecules to another. 
\end{abstract}

\maketitle

\font\smallfont=cmr7

The creation of ultracold molecules has opened up new avenues for research in physics and chemistry. 
Ultracold molecules are finding exciting applications in condensed-matter physics~\cite{zoller}, 
quantum information science~\cite{zoller1}, physics of elementary particles~\cite{epjd} 
and chemical reaction dynamics~\cite{irpc}. The experimental work on the production of dense ensembles of 
ultracold molecules is currently being pursued by many research groups~\cite{epjd}. 
The progress of the 
experiments  
relies on detailed understanding of molecular collisions at ultracold temperatures~\cite{Staanum06,Zahzam06}. 
Recent theoretical studies ~\cite{Hutson07,Quemener07,Bala97,Bala98} have provided 
important insights into atom - molecule inelastic and reactive collisions at cold and ultracold
temperatures.
The collision energy of ultracold molecules is usually much smaller than the energy 
spacing between ro-vibrational molecular levels and the duration of an ultracold collision 
is very long, which leads to unusual propensity rules and mechanisms of inelastic scattering. 
For example, vibrational relaxation in ultracold collisions of molecules with atoms  was found to give 
rise to quasi-resonant energy transfer~\cite{Forrey99} resulting in extremely narrow distributions 
over final ro-vibrational states. 
Ultracold molecules in excited vibrational states are routinely produced in experiments 
on photoassociation and Feshbach resonance linking of ultracold atoms~\cite{paul1}. 
The lifetime of the molecules is limited by collision-induced vibrational relaxation. 
Dense ensembles of ultracold molecules can also be created in experiments on buffer gas 
cooling~\cite{john} or Stark deceleration of molecular beams~\cite{gerard}. Ground-state molecules may then 
be excited by black-body radiation~\cite{gerard-bb} and release energy through vibrationally inelastic collisions. 
Laser cooling of molecules, when realized, will result in significant populations of vibrationally excited states that 
may relax through  collisions in a gas of sufficient density~\cite{dirosa}. 
Evaporative cooling of molecular ensembles to ultracold temperatures is based on energy dissipation 
in molecule - molecule collisions.
The spectrum of accessible energy levels is much denser for collisions of molecules with each other than for 
collisions of molecules with atoms and it is important to understand dominant energy transfer mechanisms 
in molecule - molecule interactions at ultracold temperatures. 

In this Letter, we present 
rigorous quantum calculations of cross sections for cold and ultracold 
collisions of p-H$_2$ molecules in different levels of vibrational and rotational excitation and identify 
a new mechanism for ro-vibrational energy transfer, not present in collisions of molecules with atoms. 
We find that the energy exchange process 
in collisions of indistinguishable molecules at cold and ultracold temperatures may be
highly efficient and state selective when the internal energy and 
total rotational angular momentum of the colliding molecules 
are simultaneously conserved. 
By treating the H$_2$ molecules as distinguishable particles it is shown that this
mechanism is primarily driven by a rotational transfer process.
We show that the efficiency of the near-resonant
energy transfer is largely insensitive to the initial
vibrational excitation of the molecules.

The H$_2$ + H$_2$ system
is the simplest neutral tetra-atomic system and it 
serves
as a prototype for describing collisions 
between diatomic molecules. 
Using a model rigid rotor potential energy surface (PES) for the H$_2$-H$_2$ system
developed by Zarur and Rabitz~\cite{Zarur74}, 
Forrey~\cite{Forrey01} investigated rotational transitions in p-H$_2$ + p-H$_2$ 
collisions at ultracold temperatures. 
Boothroyd, Martin, Keogh and Peterson (BMKP)~\cite{Boothroyd02} have recently 
calculated a full-dimensional PES for the H$_4$ system.
Pogrebnya and Clary~\cite{Pogrebnya02} reported 
rate coefficients for collisions of   
p-H$_2$ and o-H$_2$ in the first excited  vibrational state
for collision energies between 10$^{-3}$ eV (11.604 K) and 1 eV (11604 K). 
They included all internal degrees of freedom and used the BMKP PES
in a quantum calculation based on the centrifugal decoupling approximation.
Mat\'e et al.~\cite{Mate05} reported an experimental study 
of rotationally inelastic collisions at temperatures between 2~K and 110~K
and compared their data to theoretical results obtained using
the PESs calculated by Diep and Johnson (DJ)~\cite{Diep00} 
and Schaefer and K\"ohler (SK)~\cite{Schaefer89}. These potentials describe the interaction between  
H$_2$ molecules fixed at
the equilibrium internuclear distance. 
Lee et al.~\cite{Lee06} have recently presented a comparative analysis 
of cross sections for rotationally inelastic collisions between p-H$_2$
at low and ultralow energies computed using
the DJ and BMKP PESs. 
Time-dependent approaches have also been recently used to describe
the dynamics between two p-H$_2$~\cite{Lin02,Gatti05}
and between p-H$_2$ and o-H$_2$~\cite{Panda07}
at thermal energies.

The results presented here are obtained using the BMKP PES and
a full-dimensional quantum dynamics calculation
without any decoupling approximations. We employed a
newly developed quantum scattering code~\cite{Roman06} based on the formalism 
described by
Arthurs and Dalgarno~\cite{Arthurs60},
Takayanagi~\cite{Takayanagi65},
Green~\cite{Green75},
and Alexander and DePristo~\cite{Alexander77}. 
We consider H$_2$ molecules initially in different quantum states 
characterized by the vibrational quantum number $v$ and the rotational angular momentum $j$. 
We refer to a combination of two ro-vibrational states of H$_2$ as combined molecular state (CMS). 
A CMS represents a unique quantum state of the diatom - diatom system
before or after a collision.
The CMS will be denoted as $(v j v' j')$.
We used a ``well-ordered states" notation~\cite{Takayanagi65} where $v > v'$
or when $v=v'$, $j \ge j'$.
As the colliding molecules are indistinguishable,
the quantum numbers $vj$ or $v'j'$ cannot be assigned to a particular molecule.
To describe collisions of distinguishable molecules,
we will use the notation $[v j ; v' j']$, where the quantum numbers $vj$
characterize one molecule, and $v'j'$ the other.
The cross sections for state-resolved ro-vibrational transitions are obtained from the solution of coupled differential 
equations~\cite{Green75} using the log-derivative
method of Johnson~\cite{Johnson73} and Manolopoulos~\cite{Manolopoulos86},
and appropriate scattering boundary conditions~\cite{Arthurs60}.

\begin{figure} [h]
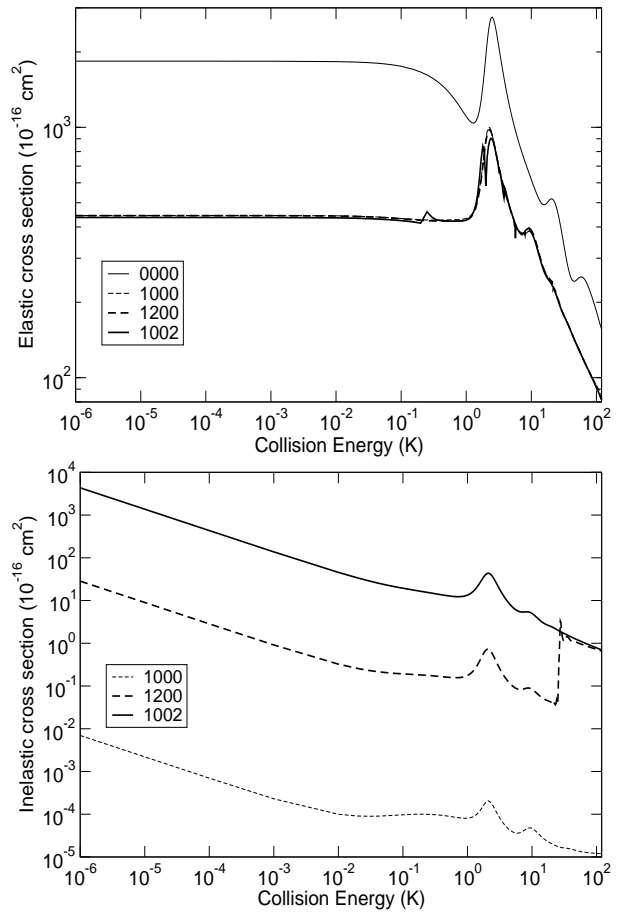

\begin{center}
\epsfig{file=fig1a.eps,height=6cm,width=8cm,clip=true}\\
\epsfig{file=fig1b.eps,height=6cm,width=8cm,clip=true}
\caption{Elastic (upper panel) and inelastic (lower panel) cross sections
for H$_2$-H$_2$ collisions for 
collision energies from 10$^{-6}$~K to 120~K. 
The curves are labeled by ($vjv'j'$) values. See text for details.
\label{XSTOT-FIG}
}
\end{center}
\end{figure}

Fig.~\ref{XSTOT-FIG} presents the cross sections for elastic and inelastic scattering of 
H$_2$ molecules for collision energies
from 10$^{-6}$~K to 120~K. 
We consider collisions of H$_2$ molecules in the excited ro-vibrational states ($v=1,j=0$) 
and $(v=1, j =2)$ with H$_2$ molecules in $(v=0, j=0)$ and collisions 
of H$_2(v=1,j=0)$ with H$_2(v=0,j=2)$.
The cross sections 
for elastic collisions of H$_2$ molecules in the ground state
$(v=0,j=0)$ are also shown.
The calculations are performed for the total angular momenta $J=[0-10]$ and for both parities. 
While the elastic scattering cross sections for collisions of molecules in the 
vibrationally excited states 
are practically the same, the probability of inelastic scattering is extremely 
sensitive to the initial states of the colliding molecules. The inelastic relaxation of H$_2(v=1,j=0)$ is six 
orders of magnitude more efficient in collisions with H$_2(v=0,j=2)$ than in collisions with H$_2(v=0,j=0)$. 
The zero-temperature limiting values of the relaxation rate constants for (1000), (1200) and (1002),
are respectively, 
$9 \times 10^{-18}$, $4 \times 10^{-14}$ and  $6 \times 10^{-12}$ cm$^3$ s$^{-1}$.
The inelastic cross section for collisions between 
H$_2(v=1,j=2)$ and H$_2(v=0,j=0)$ 
is smaller than the inelastic  cross section
for collision between
H$_2(v=1,j=0)$ and H$_2(v=0,j=2)$ at low and ultralow energies. 
It undergoes a rapid enhancement of two orders of magnitude 
at the collision energy 25.45~K, at which both inelastic cross sections become similar.

\begin{figure}[h]
\begin{center}
\epsfig{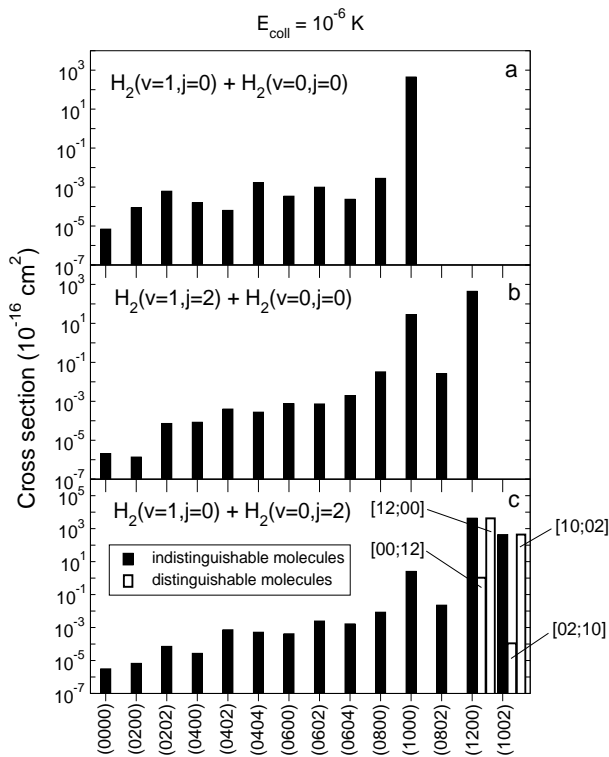}
\caption{State-to-state cross sections as functions of the final combined molecular state
for the same systems as in Fig.~\ref{XSTOT-FIG}
and for a collision energy of 10$^{-6}$~K.
\label{ROTDIST-FIG}
}
\end{center}
\end{figure}

To explain the results of Fig.~\ref{XSTOT-FIG}, 
we have calculated the corresponding state-to-state cross sections
for a collision energy of 10$^{-6}$~K (see Fig.~\ref{ROTDIST-FIG}).
The final CMSs are presented in the order of increasing energy. 
Figure~\ref{ROTDIST-FIG} shows that ro-vibrational relaxation in molecule - molecule collisions 
is determined by the propensity to conserve the internal energy and the total rotational 
angular momentum of the colliding molecules. 
The cross sections for vibrational relaxation in collisions of H$_2(v=1,j=0)$ 
with H$_2(v=0,j=0)$ 
are rather small
because 
the minimization of the energy gap between the initial and final CMS would require a large 
change of the total rotational angular momentum of the colliding molecules. 
Conversely, the conservation of the total rotational angular momentum
would introduce a large energy gap. 
Consequently, there is no single dominant CMS in the final state-to-state
distribution shown in panel $a$ of Fig.~\ref{ROTDIST-FIG}.
The inelastic relaxation of H$_2(v=1,j=2)$
in collisions with H$_2(v=0,j=0)$
shown in panel $b$ of Fig.~\ref{ROTDIST-FIG}
is dominated by the transition
$(1200) \to (1000)$. 
In this case, the energy gap and the difference in total rotational angular momentum
of the two CMSs are simultaneously minimized. 
The inelastic relaxation of H$_2(v=1,j=0)$
in collisions with H$_2(v=0,j=2)$
is however highly efficient. It is dominated by the single transition
$(1002) \to (1200)$, as shown in panel $c$ of Fig.~\ref{ROTDIST-FIG}. 
This process conserves the total rotational angular 
momentum and has a very small energy gap of 25.45~K.
The energy gap arises from the slightly different 
centrifugal distortion of the vibrational states $v=0$ and $v=1$.
Therefore, the conservation of internal energy and total rotational angular momentum
is largely satisfied and the process becomes highly efficient and selective.
This near-resonant transition is 150 times more efficient than 
the transition $(1200) \to (1000)$.
The reverse transition $(1200) \to (1002)$
is responsible for
the rapid enhancement of the inelastic cross section for H$_2(v=1,j=2)$ + H$_2(v=0,j=0)$ collisions 
at 25.45~K shown in Fig.~\ref{XSTOT-FIG}
(the state (1002) becomes energetically accessible at this collision energy). 
The mechanism
is thus a highly specific process, 
reminiscent of quasi-resonant energy transfer in collisions of 
rotationally excited diatomic molecules with atoms~\cite{Stewart98,Forrey99,heller06},
but with a purely quantum origin.

In order to  
elucidate the mechanism for the (1002) $\to$  (1200) transition, we 
have repeated the calculation in Fig.~\ref{ROTDIST-FIG} 
for the collision process H$_2(v=1,j=0)$ + H$_2(v=0,j=2)$ 
assuming that the colliding molecules are distinguishable.
The inelastic transition (1002) $\to$ (1200)
in collisions of indistinguishable molecules
corresponds to two transitions: [10;02] $\to$ [12;00]
and [10;02] $\to$ [00;12]
in collisions of distinguishable molecules.
The first one 
corresponds to purely rotational energy transfer,
and the second to purely vibrational energy transfer. 
As follows from Fig.~\ref{ROTDIST-FIG}, 
the rotational energy transfer
dominates over 
the vibrational energy transfer by 
three orders of magnitude.
The inelastic transition (1002) $\to$  (1200)
can thus be viewed
mainly as the rotational energy transfer.
These symmetry considerations 
can also explain why Panda et al.~\cite{Panda07}
found a surprisingly small cross section for 
the resonant distinguishable
process p-H$_2(v=0,j=0)$ + o-H$_2(v=1,j=1)$ $\to$ 
p-H$_2(v=1,j=0)$ + o-H$_2(v=0,j=1)$, 
despite a small energy gap of 8.5~K.
The rotational energy transfer between $j=0$ and $j=1$
is forbidden due to symmetry
so this transition is driven by the vibrational energy transfer.
We conclude that 
the near-resonant transitions should be generally smaller 
in ortho-para collisions
than in para-para collisions.

\begin{figure}[h]
\begin{center}
\epsfig{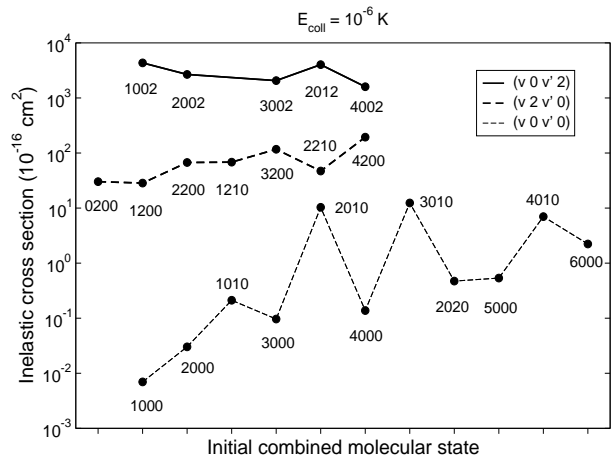}
\caption{Inelastic cross sections
as functions of the initial combined molecular state
for a collision energy of $10^{-6}$~K.
\label{XS-allV-FIG}
}
\end{center}
\end{figure}

Fig.~\ref{XS-allV-FIG} demonstrates the effects of vibrational excitation of the colliding molecules 
on the inelastic relaxation cross section at a collision energy of $10^{-6}$~K.
The results are presented as functions of the initial
CMSs $(v0v'0)$ (dashed line), $(v2v'0)$ (dashed bold line) and $(v0v'2)$ (solid line).
We use the full basis set for the calculation involving the initial CMSs
$(0j0j')$, $(1j0j')$, $(2j0j')$ and $(1j1j')$ and 
truncated the basis set, as described in Ref.~\cite{romanHeCO}, for the calculations focusing on 
higher vibrational states. 
The truncation eliminates all channels corresponding to
$\Delta v \text{ and/or } \Delta v' > 1$ transitions. 
The magnitude of the inelastic cross sections for collisions of H$_2(v,j=0)$ with H$_2(v',j=0)$ 
depends strongly on the initial vibrational states of the colliding molecules
which is consistent with the previous analysis of vibrational relaxation in
H-H$_2$~\cite{Bala97}, He-H$_2$~\cite{Bala98}, and  Li-Li$_2$~\cite{Quemener07}
collisions.
For $v=[1-6]$ and $v'=0$, the cross sections show a monotonous increase  with
the vibrational quantum number $v$
as observed for He-H$_2$ collisions~\cite{Bala98}
at ultralow energies.
When both molecules are initially vibrationally excited as
in $v=[2-4]$ and $v'=1$, the inelastic cross sections become significantly larger
and less sensitive to the initial vibrational levels of the two molecules.
For collisions of H$_2(v,j=2)$ with H$_2(v',j=0)$, the inelastic cross sections 
are quite large and show a slight increase with the vibrational quantum numbers. 
Finally, for collisions of H$_2(v,j=0)$ with H$_2(v',j=2)$, the inelastic cross sections
are very large due to the near-resonant energy transfer and do not display 
any particular dependence on the vibrational levels. 
We have found that the elastic cross sections generally do not depend 
strongly on the vibrational excitation
of the molecules. They rather depend on quantum statistics effect, whether or not
the vibrational and rotational quantum numbers of the colliding molecules are the same~\cite{Biolsi73}. 
This difference can be seen in the elastic cross sections 
of (0000) and (1000) in Fig.~\ref{XSTOT-FIG}.

In summary, we have presented the first study of vibrational relaxation in cold and
ultracold molecule - molecule collisions and found a new mechanism of state selective
ro-vibrational energy transfer. 
If the colliding molecules are in different levels of vibrational and rotational excitation, 
the ro-vibrational energy may be efficiently exchanged between the molecules,
leading to near-resonant energy transfer. If the molecules
are distinguishable, this process corresponds to rotational
energy transfer from one vibrational state of the colliding molecules to 
another. This inelastic process that dominates over all other inelastic 
transitions is largely insensitive to the initial vibrational excitation 
of the molecules. 
The process may be an important mechanism for collisional energy 
transfer in ultracold molecules formed by photoassociation of ultracold
atoms and for chemical reactions producing identical
molecules. We hope that these results will stimulate further studies of 
vibrational energy transfer in molecular collisions at ultracold 
temperatures.

This work was supported  by NSF grant \# PHY-0555565 (N.B), Natural Sciences and Engineering 
Council (NSERC) of Canada (R.V.K.), NASA Grant  NNG06GJ11G from the Astrophysics Theory Program 
and NASA grant \# NNG06GC94G to David Huestis of SRI International.

\end{document}